\documentstyle[axodraw]{article}

\def\ltap{\raisebox{-.4ex}{\rlap{$\sim$}} \raisebox{.4ex}{$<$}}

\newcommand{\Rsl}{{\not \! \!{R}}}
\newcommand{\sqm}{m_{\tilde{q}}}
\newcommand{\slm}{m_{\tilde{l}}}

\begin{document}
\vspace*{-1in}
\renewcommand{\thefootnote}{\fnsymbol{footnote}}
\begin{flushright}
SINP/TNP/98-31\\
CUPP--98/2 \\
\texttt{hep-ph/9811500} 
\end{flushright}
\vskip 5pt
\begin{center}
{\Large{\bf $R$-parity-violating trilinear
couplings and recent neutrino data}}
\vskip 25pt
{\sf Subhendu Rakshit $^{a,\!\!}$
\footnote{E-mail address: srakshit@cucc.ernet.in}}, 
{\sf Gautam Bhattacharyya $^{b,\!\!}$
\footnote{E-mail address: gb@tnp.saha.ernet.in}}   
and 
{\sf Amitava Raychaudhuri $^{a,\!\!}$
\footnote{E-mail address: amitava@cubmb.ernet.in}}  
\vskip 10pt
$^a${\it Department of Physics, University of Calcutta, 
92 Acharya Prafulla Chandra Road, \\ Calcutta 700009, India} \\
$^b${\it Saha Institute of Nuclear Physics, 1/AF Bidhan Nagar, Calcutta 
700064, India}
\vskip 20pt

{\bf Abstract}
\end{center}

\begin{quotation}
{\small The nontrivial structure of the neutrino mass matrix,
suggested by the recent Super-Kamiokande results and data from other
neutrino experiments, can be reproduced in $R$-parity-violating
supersymmetric theories. This requires sets of products of
$R$-parity-violating trilinear couplings to take appropriately chosen
values. It is shown that the existing constraints on these couplings
are satisfied by these choices.
\\ PACS number(s): 12.60.J, 11.30.Fs, 14.60.Pq}
\end{quotation}

\vskip 20pt  

\setcounter{footnote}{0}
\renewcommand{\thefootnote}{\arabic{footnote}}
%\vfill
%\clearpage
%\setcounter{page}{1}
%\pagestyle{plain}

The recent data from the Super-Kamiokande (SK) collaboration provides
strong evidence in support of neutrino oscillations as an explanation
of the atmospheric anomaly \cite{superk}. The observed solar neutrino
deficit \cite{solar} and the LSND accelerator experiment \cite{lsnd}
are also indicative of neutrino oscillations. Put together, all these
evidences indicate a nontrivial structure of the neutrino mass
matrix. In this paper we show that this mass matrix can be reproduced
in the $R$-parity violating ($\Rsl$) Minimal Supersymmetric Standard
Model \cite{rpar} and requires sets of products of the couplings to
take up appropriate values which are consistent with existing
constraints.

`R-parity' in supersymmetry refers to a discrete symmetry which
follows from the conservation of lepton-number ($L$) and baryon-number
($B$). It is defined as $R=(-1)^{(3B+L+2S)}$, where $S$ is the
intrinsic spin of the field. $R$ is $+1$ for all standard model
particles and $-1$ for all super-particles.  However, $B$- and $L$-
conservation are not ensured by gauge invariance and hence there is
{\it a priori} no reason to set these couplings to zero.

The most general $\Rsl$ superpotential is given by,
\begin{equation}
 W_{\Rsl} 
={1 \over 2} \lambda_{ijk} L_{i} L_{j} E^{c}_{k}
           + \lambda'_{ijk} L_{i} Q_{j} D^{c}_{k}
          + {1 \over 2} \lambda''_{ijk} U^{c}_{i} D^{c}_{j} D^{c}_{k}
           +\mu_{i} L_{i} H_u, 
\label{superpot}
\end{equation}
where $i, j, k = 1, 2, 3$ are quark and lepton generation indices;
$L_i$ and $Q_i$ are $SU(2)$-doublet lepton and quark superfields
respectively; $E^{c}_i$, $U^{c}_i$, $D^{c}_i$ are $SU(2)$-singlet
charged lepton, up- and down-type quark superfields respectively;
$H_u$ is the Higgs superfield responsible for the generation of
up-type quark masses; $\lambda_{ijk}$ and $\lambda'_{ijk}$ are
$L$-violating while $\lambda''_{ijk}$ are $B$-violating Yukawa
couplings.  $\lambda_{ijk}$ is antisymmetric under the interchange of
the first two generation indices, while $\lambda''_{ijk}$ is
antisymmetric under the interchange of the last two. Thus there could
be 27 $\lambda'$, 9 each of $\lambda$ and $\lambda''$ couplings and 3
$\mu_i$ parameters. We assume that the generation indices correspond
to the flavour basis of fermions. We note at this point that proton
stability severely restricts the upper limits on the products of $B$-
and $L$-violating couplings \cite{pdk}. Therefore our requirement of
`not-too-small' $L$-violating couplings for the present analysis
implies that the $B$-violating couplings are either zero or
vanishingly small. In any case, the $\lambda''_{ijk}$ couplings are
not of any relevance to the present work.

Stringent constraints on individual $L$-violating couplings have been
placed from the consideration of neutrinoless double beta decay,
$\nu_e$-Majorana mass, charged-current universality, $e-\mu-\tau$
universality, $\nu_{\mu}$ deep-inelastic scattering, atomic parity
violation, $\tau$ decays, $D$ and $K$ decays, $Z$ decays, etc. Product
couplings (two at a time), on the other hand, have been constrained by
considering $\mu-e$ conversion, $\mu\rightarrow e\gamma$,
$b\rightarrow s \gamma$, $B$ decays into two charged leptons,
$K_L-K_S$ and $B_q-\bar{B_q}$ ($q = d,s$) mass differences, etc. (For
a collection of all these limits, see~\cite{review}).

In this work, we are concerned with the generation of neutrino masses
and mixings.  One of the neutrino states can develop a non-zero
tree-level mass from the last term of eq.~(\ref{superpot}). This
obtains if $\mu_\alpha$ and the vacuum expectation values $\langle
L_\alpha\rangle$, where $\alpha = 0,..,3$, considered as two four
component vectors, are not aligned \cite{misal}. Here, $L_0 \equiv
H_d$ (the Higgs responsible for the mass generation of down-type
quarks and charged leptons) and $\mu_0 \equiv \mu$ (note, $\mu H_d
H_u$ appears in the $R$-parity-conserving superpotential). This
procedure cannot generate the other terms of the mass matrix. Here we
are interested in reproducing the complete mass matrix as masses and
mixings of neutrinos are both essential ingredients of our analysis.
So we concentrate on the alternative way in which Majorana masses are
generated at 1-loop order {\em via} self-energy diagrams involving
$\lambda$ or $\lambda'$ couplings (we will discuss this in detail in
the following paragraphs). The complete mass matrix can be generated
by considering different leptonic flavour indices attached to
$\lambda$ or $\lambda'$. For the sake of simplicity, we choose the
$\mu_i$ terms in~(\ref{superpot}) to be zero.  Although this amounts
to a slight loss of generality, dropping these terms helps to examine
in isolation the r\^{o}le of the trilinear couplings, the focus of our
analysis, in reproducing the recent neutrino data. Moreover, the
numerical impact of the tree-level contribution induced by the $\mu_i$
terms could be tuned to be very small by arranging a perfect or a
close alignment between $\mu_\alpha$ and $\langle L_\alpha\rangle$ at
some high scale. The misalignment that might creep in through
renormalisation group running is suppressed by loop factors \cite
{CWP,chun}. As a result, it is possible to arrange that the other
contribution, namely, the one induced by trilinear $L$-violating
terms, dominates. Hence for the order of magnitude estimate of the
values that we assign on the product of trilinear couplings, dropping
the $\mu_i$ parameters is not unjustified. We are also not interested
in see-saw type contributions to neutrino masses, which involve heavy
right-handed neutrinos.

Majorana mass terms for the left-handed neutrinos can be
generated through quark-squark loop diagrams (Fig.~1(a))
which involve the $\lambda'$ couplings in the following way:
\begin{equation}
m_{\nu_{ii'}} \approx {{3} \over{8\pi^2}} \lambda'_{ijk}
\lambda'_{i'kj} {{m_{d_j} \Delta m_k^2(d)} \over {\sqm}^2}.
\label{lplp}
\end{equation}
These mass terms can also be generated through
lepton-slepton loop diagrams (Fig.~1(b)) which are related 
to the $\lambda$ couplings as:
\begin{equation}
m_{\nu_{ii'}} \approx {{1} \over{8\pi^2}} \lambda_{ijk}
\lambda_{i'kj} {{m_{e_j} \Delta m_k^2(l)} \over {\slm}^2}.
\label{ll}
\end{equation}
In eqs.~(\ref{lplp}) and (\ref{ll}), $\Delta m^2 (f)$ represents the
left-right sfermion mixing term which we assume can be parametrized as
$\Delta m^2 (f) \approx m_f \tilde{m}$, where $\tilde{m}$ is the
average squark mass $\sqm$ and the average slepton mass $\slm$ in
eqs.~(\ref{lplp}) and (\ref{ll}) respectively.

Strictly speaking, in eq.~(\ref{lplp}) some quark mixing angles ought
to appear. This is because the $\Rsl$-interactions are written in the
flavour basis while the states propagating in the loop diagrams are
the mass eigenstates. If the entire Cabibbo-Kobayashi-Maskawa mixing
is attributed to the down-type quark sector then the replacement
$m_{d_j} \rightarrow \Sigma_l |V_{lj}|^2 m_{d_l}$ is needed. On the
contrary, if the entire quark mixing is in the up-type sector then no
changes are necessary. It also needs to be remarked that at the two
vertices d-type quarks of opposite chiralities appear. The mixing of
quarks in the right-handed sector cannot be probed {\em via} the
Standard Model interactions and the choice which yields the factor
mentioned earlier corresponds to taking identical mixing for both
chiralities. In order not to complicate matters unnecessarilly, we
have ignored this small difference in the flavour and mass bases in
the following.

The mass matrices generated in this way correspond to the
flavour-basis of fermions. The expressions above resemble the see-saw
$m_D^2/M$ formula to some extent but there are several
differences. $m_D$ is the neutrino Dirac mass -- in GUT motivated
models related to the up-type quark mass -- whereas here we have
charged lepton or down-type quark masses in the numerator. More
importantly, the smallness of the neutrino masses in this picture is
due not just to mass ratios but also to the sizes of the $\Rsl$
interactions.

We consider mixing of the three neutrinos as:
\begin{equation}  
\pmatrix{\nu_e \cr \nu_{\mu} \cr \nu_{\tau}} = \cal{U} 
\pmatrix{\nu_1 \cr \nu_2 \cr \nu_3}
\end{equation} 
where $\nu_e, \nu_{\mu}$ and $\nu_{\tau}$ are the flavour states and
$\nu_1, \nu_2$ and $\nu_3$ are mass eigenstates with masses
$m_{\nu_1}, m_{\nu_2}$ and $m_{\nu_3}$ respectively, which we choose
to satisfy the hierarchical mass structure
$m_{\nu_1}<m_{\nu_2}<m_{\nu_3}$. The mixing matrix $\cal{U}$ is taken
to be real and can be parametrised in terms of three mixing angles
$\theta_{12}, \theta_{23}$ and $\theta_{13}$ as

\begin{equation} 
{\cal{U}} = 
\pmatrix{c_{12} c_{13} & s_{12} c_{13} & s_{13} 
       \cr -s_{12} c_{23} -c_{12} s_{23} s_{13} & c_{12} c_{23}
-s_{12} s_{23} s_{13} & s_{23} c_{13} \cr s_{12} s_{23} - c_{12}
c_{23} s_{13} & -c_{12} s_{23} -s_{12} c_{23} s_{13} & c_{23}
c_{13}}
\end{equation} 
where $c$ and $s$ stand for cosine and sine respectively of the
mixing angles.

As is well known, neutrino oscillations depend on two factors: (a) the
flavour eigenstates and mass eigenstates do not coincide, and (b) the
mass eigenstates are not degenerate. The experimental indications of
neutrino oscillations can be used to restrict the possible ranges of
the mixing angles and neutrino mass splittings, $\Delta m^2$.  For
example, it has been shown~\cite{fogli} that the recent SK data, along
with the CHOOZ~\cite{chooz} and the pre-SK solar neutrino results, can
be accommodated in a three neutrino oscillation model at 99\% CL, with
$\sin^2 \theta_{12}=0.4$, $\sin^2 \theta_{23}=0.5$, $\sin^2
\theta_{13}=0.2$, $\Delta m^2_{32} =m_{\nu_{3}}^2-m_{\nu_{2}}^2=8
\times 10^{-4}~{\rm eV}^2$, $\Delta
m^2_{21}=m_{\nu_{2}}^2-m_{\nu_{1}}^2=1 \times 10^{-4}~{\rm
eV}^2$. Another group \cite{thun} claimed a good overall fit to the
atmospheric SK data for $\theta_{12}=37.6^o$, $\theta_{23}=26.5^o$,
$\theta_{13}=10.3^o$, $\Delta m^2_{32}=0.4~{\rm eV}^2$, $\Delta
m^2_{21}=0.0003~{\rm eV}^2$, including the LSND and solar neutrino
results in their analysis.  A similar solution is also offered in
ref.~\cite{baren}. It needs to be mentioned that questions about the
result in ref.~\cite{thun} have been raised in ref.~\cite{fogli}
partly because matter effects were ignored in the analysis of
ref.~\cite{thun}. Further, the data from the Homestake experiment were
also ignored in ref.~\cite{thun}. We have examined the requirements
for the values of $\Rsl$ couplings from {\em both} these
fits~\cite{fogli,thun} and found that those implied by the results in
ref.~\cite{fogli} are more stringent than the other one. A similar
effort, but restricted to only some individual $\Rsl$ couplings and
not products, has been presented in~\cite{rathin} using the results of
\cite{thun}. We agree with the results of ref.~\cite{rathin} in the
appropriate limits. An analysis containing only $\lambda$-type
couplings along with a neutrino mass hierarchy inverted with respect
to ours has been presented in ref.~\cite{clav}.

Once the mixing matrix for the neutrinos and the mass splittings
are fixed, the structure of the neutrino mass matrix is
determined. Choosing any one of the neutrino masses completely
specifies this matrix.  Of course, one must bear in mind that
there are some experimental limits which must be respected.  The
(11) component of the Majorana mass matrix in the flavour basis
is constrained to be less than $0.46$ eV from neutrinoless double
beta decay experiments~\cite{baudis}. The masses of $\nu_{\mu}$
and $\nu_{\tau}$ are constrained to be $m_{\nu_{\mu}} \leq 0.17$
MeV~\cite{pdg} and $m_{\nu_{\tau}} \leq 18.2$ MeV~\cite{pdg}.

Here we have performed our analysis for $m_{\nu_1} =$ 0, 0.01, and 0.1
eV taking $\slm = \sqm =$ 100 GeV. In Table 1 we present the values
which the products of $\lambda'$-type couplings must assume in order
to reproduce the neutrino mass matrix. Note that each element of the
mass matrix can be generated by several different product
couplings. It should be borne in mind that the presented values are
{\it not} upper bounds. The complete mass matrix must be reproduced in
order to obtain the correct mass eigenvalues and eigenstates and for
each element of the mass matrix any one of the corresponding product
couplings listed in Table 1 must achieve the corresponding listed
value. It is noteworthy that in several of the cases presented in
Table 1 there are strong existing constraints from other processes on
the relevant product couplings forbidding the required value. It may
be mentioned that in case the sparticle masses and mixings are
determined from some model ({\em e.g.} supergravity) then the specific
values of the product couplings that we have presented will be
replaced by ranges dependent on the model parameters ({\em e.g.}
$\mu,\tan\beta$ etc.).  Further, in obtaining the results presented in
Table 1 we have used as input a set of mass-splittings and mixing
angles which accommodates the data at 99\% CL \cite{fogli}. This is
not a best-fit set and, indeed, at 99\% CL for each of the parameters
there will be an allowed range\footnote{Such a range is unavailable in
the published literature. Nevertheless, the given mass differences and
mixing angles enable us to examine our primary concern in this paper,
{\em i.e.}  whether the $\Rsl$ Yukawa couplings, with assigned values
allowed otherwise, are indeed capable of reproducing the observed
indication of neutrino masses and mixings.}. These will, in turn, lead
to allowed ranges for the product couplings rather than the specific
values listed in Table 1.

Table 2 is a similar list but for products of the $\lambda$-type
couplings. For ease of presentation, in this Table we have not
followed an oft-used convention in which for $\lambda_{ijk}$ the
antisymmetry in $i$ and $j$ is utilised to always choose $i < j$.
Note that due to the antisymmetry, this Table has fewer entries
than the previous one.

A somewhat similar analysis, in a two-generation
$\nu_{\mu}-\nu_{\tau}$ oscillation scenario, has been performed in
ref.~\cite{kong}. Assuming $\Delta m^2_{\nu_{\mu} \nu_{\tau}} \simeq
m_{\nu_{\tau}}^2$, it has been shown that $\lambda'_{233}$ (also,
$\lambda'_{333}$) $\sim 10^{-5}$ and $\lambda_{233}$ (also,
$\lambda_{232}$) $\sim 10^{-4}$ are relevant for explaining the
atmospheric neutrino oscillation anomaly. Our analysis is based on
three-generation neutrino oscillation and we do not consider the
effects of $\mu_i$ terms. On account of mainly these two differences,
the values of the $\Rsl$ product couplings that we have found are a
little larger in comparison with those of ref.~\cite{kong}.

At this point, a discussion of the results of Tables 1 and 2 are in
order.  We observe that as $m_{\nu_1}$ increases, the magnitudes of
the $\lambda'_{ijk} \lambda'_{ikj}$ type products increase, whereas
$\lambda'_{ijk} \lambda'_{i'kj}(i \neq i')$ type products behave
oppositely. The same is true for the products of the $\lambda$
couplings as well. This fact is easily comprehensible in a two
generation scenario. Notice that the values of the product couplings
which we find get diluted with increased scalar masses as
$\tilde{m}/100$ GeV, where $\tilde{m}$ is the mass of the relevant
scalar. Here we list the dependences of the existing bounds for some
of the individual couplings (see, ref.~\cite{review}) on the squark or
slepton masses: $\lambda'_{111} \sim
({m_{\tilde{u}_{L},\tilde{d}_{R}}}/{100~{\rm GeV}})^2$,
$\lambda'_{11k(k \neq 1)} \sim ({m_{\tilde{d}_{kR}}}/ 100~{\rm GeV})$,
$\lambda'_{1j1(j \neq 1)} \sim ({m_{\tilde{q}_{jL}}}/ 100~{\rm GeV})$,
$\lambda'_{1jj} \sim ({m_{\tilde{d}_j}}/ 100~{\rm GeV})^{1/2}$,
$\lambda'_{21k} \sim ({m_{\tilde{d}_{kR}}}/ 100~{\rm GeV})$,
$\lambda'_{231} \sim ({m_{\tilde{\nu}_{\tau L}}}/ 100~{\rm GeV})$,
$\lambda_{133} \sim ({m_{\tilde{\tau}}}/ 100~{\rm GeV})^{1/2}$ and all
other $\lambda_{ijk} \sim ({m_{\tilde{e}_{kR}}}/ 100~{\rm GeV})$. For
$\lambda'_{132}$, $\lambda'_{22k}$, $\lambda'_{23k(k \neq 1)}$,
$\lambda'_{31k}$ and $\lambda'_{33k}$ the dependences are more
complicated. For $\lambda'_{i12} \lambda'_{i21}$-, and $\lambda'_{i13}
\lambda'_{i31}$-combinations, on the other hand, it is not necessary
to take products of individual couplings.  These are constrained from
tree level $\Delta S = 2$ and $\Delta B = 2$ processes
respectively. The bounds are $\lambda'_{i12} \lambda'_{i21} \ltap 1
\times 10^{-9}$ $(m_{\tilde{\nu}_L}/ 100~{\rm GeV})^2$ and
$\lambda'_{i13} \lambda'_{i31} \ltap 8 \times 10^{-8}$
$(m_{\tilde{\nu}_L}/ 100~{\rm GeV})^2$~\cite{review}. As listed in the
Tables, the product couplings relevant for our studies of the neutrino
mass matrix are bounded from their contributions to other physical
processes. It would not be out of place to stress here that the
processes from which these bounds are obtained involve exchanged
scalars which are not the same and their masses are generally
uncorrelated. Notice that we have followed the usual practice of
comparing the magnitudes of product couplings from different processes
assuming a benchmark value of 100 GeV for whichever scalar is
involved. Needless to say, such a comparison should be made in a
guarded way since these scalars could be highly non-degenerate. As an
example, if one considers $R$-parity breaking in gauge-mediated
supersymmetry breaking models \cite{GR}, where squarks are much
heavier than sleptons on account of the former's strong coupling
dependence in comparison with the latter's weak, all bounds which
depend on squark masses become significantly weaker as one cannot
admit a squark mass as low as 100 GeV in the phenomenological
description of such models.

We make a remark in passing that a heavier neutrino state can in
principle decay radiatively into a lighter state {\em via} graphs
involving trilinear $L$-violating couplings. Such decays are
cosmologically troublesome \cite{GeRou}. However, on account of the
low mass (order eV) of even the heaviest state and the smallness of
the couplings, the decay proceeds at an extremely slow rate so that it
does not take place within the present lifetime of the universe (order
$10^{18}$ s). Finally, we point out that although quite a few non-zero
$L$-violating couplings are required to be present simultaneously to
reproduce the complete mass matrix in our analysis, they do not, with
the kind of numbers they need to assume, trigger any forbidden or
highly suppressed process at an unwanted rate. It has to be mentioned
that if one attempts to realise $R$-parity as an extension of the
conventional flavour problem in supersymmetry in the context of an
abelian \cite{u1} or a non-abelian \cite{gb} flavour group,
many non-zero $\Rsl$-couplings naturally appear together with
appropriate suppressions dictated by the flavour symmetry.

In conclusion, we have shown that in supersymmetry it is possible to
reproduce the neutrino mass matrix, as determined by the latest
experimental data, through loop diagrams which involve products of
$\Rsl$ trilinear couplings provided sets of these products take on
specific values. All the existing constraints on these parameters are
satisfied. Indeed, this might not be the only source of neutrino mass
generation of which there is a wide latitude of possibilities existing
in the literature. Still our mechanism is an `existence proof' in
support of the observed data. Our effort to relate the smallness of
neutrino masses to the smallness of the $\Rsl$ Yukawa couplings might
provide an indication of their common ancestral link rooted to some
underlying flavour theory.

\vskip 10pt \noindent 
SR acknowledges support from the Council of Scientific and Industrial
Research, India. AR has been supported in part by the Council of
Scientific and Industrial Research and the Department of Science and
Technology, India.

%\newpage

\newpage

%\begin{center} 
{\sf \noindent Table 1: Values ({\em not} bounds) of the
$\lambda'$-type product couplings for different possible $\nu_1$
masses. In deriving our numbers, we have used the results of
ref.~\cite{fogli}. The products marked with `*' are obtained by
multiplying the upper bounds on the individual couplings.}
$$
\begin{array}{|c|c|c|c|c|c|c|} 
\hline
{\rm Mass ~ matrix} & {\rm Combinations} &  m_{\nu_1}=0~{\rm eV} 
& m_{\nu_1}=0.01~{\rm eV} & m_{\nu_1}=0.1~{\rm eV} 
& {\rm Existing ~ bounds} \\
{\rm elements} & & & & & \\
\hline
& \lambda'_{111} \lambda'_{111} & 9.7~~ 10^{-4} & 1.6~~ 10^{-3} 
& 1.1~~ 10^{-2} & 1.2~~ 10^{-7}  \cite{review}^*\\
& \lambda'_{112} \lambda'_{121} & 4.8~~ 10^{-5} & 8.2~~ 10^{-5} 
& 5.3~~ 10^{-4} & 1~~ 10^{-9} \cite{review} \\ 
M_{11} 
& \lambda'_{113} \lambda'_{131} & 1.1~~ 10^{-6} & 1.8~~ 10^{-6} 
& 1.2~~ 10^{-5} & 8~~ 10^{-8} \cite{review} \\ 
& \lambda'_{122} \lambda'_{122} & 2.4~~ 10^{-6} & 4.1~~ 10^{-6} 
& 2.7~~ 10^{-5} & 4~~ 10^{-4}  \cite{review}^*\\ 
& \lambda'_{123} \lambda'_{132} & 5.3~~ 10^{-8} & 9.0~~ 10^{-8} 
& 5.8~~ 10^{-7} & 1.4~~ 10^{-2} \cite{review,web}^* \\ 
& \lambda'_{133} \lambda'_{133} & 1.1~~ 10^{-9} & 2.0~~ 10^{-9} 
& 1.3~~ 10^{-8} &  4.9~~ 10^{-7} \cite{review}^*\\
\hline
& \lambda'_{111} \lambda'_{211} & 1.1~~ 10^{-3} & 7.3~~ 10^{-4} 
& 1.4~~ 10^{-4} & 5~~ 10^{-8}  \cite{review}\\ 
& \lambda'_{112} \lambda'_{221} & 5.5~~ 10^{-5} & 3.7~~ 10^{-5} 
& 7.1~~ 10^{-6} & 3.6~~ 10^{-3}  \cite{review}^*\\ 
& \lambda'_{113} \lambda'_{231} & 1.2~~ 10^{-6} & 7.9~~ 10^{-7} 
& 1.5~~ 10^{-7} & 4.4~~ 10^{-3} \cite{review}^* \\ 
& \lambda'_{121} \lambda'_{212} & 5.5~~ 10^{-5} & 3.7~~ 10^{-5} 
& 7.1~~ 10^{-6} & 3.2~~ 10^{-3} \cite{review}^* \\ 
M_{12} = M_{21} 
& \lambda'_{122} \lambda'_{222} & 2.8~~ 10^{-6} & 1.8~~ 10^{-6} 
& 3.5~~ 10^{-7} & 3.6~~ 10^{-3} \cite{review}^* \\ 
& \lambda'_{123} \lambda'_{232} & 6.0~~ 10^{-8} & 4.0~~ 10^{-8} 
& 7.7~~ 10^{-9} & 1.4~~ 10^{-2} \cite{review,web}^* \\ 
& \lambda'_{131} \lambda'_{213} & 1.2~~ 10^{-6} & 7.9~~ 10^{-7} 
& 1.5~~ 10^{-7} & 3.2~~ 10^{-3} \cite{review}^* \\ 
& \lambda'_{132} \lambda'_{223} & 6.0~~ 10^{-8} & 4.0~~ 10^{-8} 
& 7.7~~ 10^{-9} & 6.1~~ 10^{-2} \cite{review}^* \\ 
& \lambda'_{133} \lambda'_{233} & 1.3~~ 10^{-9} & 8.6~~ 10^{-10} 
& 1.7~~ 10^{-10} & 2.5~~ 10^{-4} \cite{review}^* \\ 
\hline
& \lambda'_{111} \lambda'_{311} & 4.5~~ 10^{-4} & 4.6~~ 10^{-4} 
& 1.1~~ 10^{-4} & 3.5~~ 10^{-5} \cite{review}^* \\ 
& \lambda'_{112} \lambda'_{321} & 2.2~~ 10^{-5} & 2.3~~ 10^{-5} 
& 5.4~~ 10^{-6} & 7.2~~ 10^{-3}\cite{review,web}^* \\ 
& \lambda'_{113} \lambda'_{331} & 4.9~~ 10^{-7} & 5.0~~ 10^{-7} 
& 1.2~~ 10^{-7} & 9.6~~ 10^{-3} \cite{review}^* \\ 
& \lambda'_{121} \lambda'_{312} & 2.2~~ 10^{-5} & 2.3~~ 10^{-5} 
& 5.4~~ 10^{-6} & 3.5~~ 10^{-3} \cite{review}^* \\ 
M_{13} = M_{31} 
& \lambda'_{122} \lambda'_{322} & 1.1~~ 10^{-6} & 1.2~~ 10^{-6} 
& 2.7~~ 10^{-7} & 7.2~~ 10^{-3}\cite{review,web}^* \\ 
& \lambda'_{123} \lambda'_{332} & 2.4~~ 10^{-8} & 2.5~~ 10^{-8} 
& 5.9~~ 10^{-9} & 1.9~~ 10^{-2} \cite{review,web}^* \\ 
& \lambda'_{131} \lambda'_{313} & 4.9~~ 10^{-7} & 5.0~~ 10^{-7} 
& 1.2~~ 10^{-7} & 3.5~~ 10^{-3} \cite{review}^* \\ 
& \lambda'_{132} \lambda'_{323} & 2.4~~ 10^{-8} & 2.5~~ 10^{-8} 
& 5.9~~ 10^{-9} & 1.4~~ 10^{-1}\cite{review,web}^* \\ 
& \lambda'_{133} \lambda'_{333} & 5.3~~ 10^{-10} & 5.4~~ 10^{-10} 
& 1.3~~ 10^{-10}  & 3.4~~ 10^{-4} \cite{review}^* \\ 
\hline
& \lambda'_{211} \lambda'_{211} & 1.4~~ 10^{-3} & 2.0~~ 10^{-3} 
& 1.1~~ 10^{-2} & 8.1~~ 10^{-3} \cite{review}^*\\ 
& \lambda'_{212} \lambda'_{221} & 7.0~~ 10^{-5} & 1.0~~ 10^{-4} 
& 5.4~~ 10^{-4} & 1~~ 10^{-9} \cite{review} \\ 
M_{22}
& \lambda'_{213} \lambda'_{231} & 1.5~~ 10^{-6} & 2.2~~ 10^{-6} 
& 1.2~~ 10^{-5} & 8~~ 10^{-8} \cite{review} \\ 
& \lambda'_{222} \lambda'_{222} & 3.5~~ 10^{-6} & 5.0~~ 10^{-6} 
& 2.7~~ 10^{-5} &  3.2~~ 10^{-2} \cite{review}^*\\ 
& \lambda'_{223} \lambda'_{232} & 7.6~~ 10^{-8} & 1.1~~ 10^{-7} 
& 5.8~~ 10^{-7} &  6.5~~ 10^{-2} \cite{review}^*\\ 
& \lambda'_{233} \lambda'_{233} & 1.6~~ 10^{-9} & 2.4~~ 10^{-9} 
& 1.3~~ 10^{-8} &  1.3~~ 10^{-1} \cite{review}^*\\             
\hline
& \lambda'_{211} \lambda'_{311} & 9.9~~ 10^{-4} & 8.0~~ 10^{-4} 
& 1.7~~ 10^{-4} & 9~~ 10^{-3} \cite{review}^* \\ 
& \lambda'_{212} \lambda'_{321} & 5.0~~ 10^{-5} & 4.0~~ 10^{-5} 
& 8.6~~ 10^{-6} & 1.8~~ 10^{-2} \cite{review,web}^* \\ 
& \lambda'_{213} \lambda'_{331} & 1.1~~ 10^{-6} & 8.7~~ 10^{-7} 
& 1.9~~ 10^{-7} & 4.3~~ 10^{-2} \cite{review}^* \\ 
& \lambda'_{221} \lambda'_{312} & 5.0~~ 10^{-5} & 4.0~~ 10^{-5} 
& 8.6~~ 10^{-6} & 1.8~~ 10^{-2} \cite{review}^*\\ 
M_{23} = M_{32} 
& \lambda'_{222} \lambda'_{322} & 2.5~~ 10^{-6} & 2.0~~ 10^{-6} 
& 4.3~~ 10^{-7} & 2.2~~ 10^{-1} \cite{review,web}^* \\ 
& \lambda'_{223} \lambda'_{332} & 5.4~~ 10^{-8} & 4.3~~ 10^{-8} 
& 9.3~~ 10^{-9} & 8.6~~ 10^{-2} \cite{review}^* \\ 
& \lambda'_{231} \lambda'_{313} & 1.1~~ 10^{-6} & 8.7~~ 10^{-7} 
& 1.9~~ 10^{-7} & 2.2~~ 10^{-2} \cite{review}^* \\ 
& \lambda'_{232} \lambda'_{323} & 5.4~~ 10^{-8} & 4.3~~ 10^{-8} 
& 9.3~~ 10^{-9} & 1.3~~ 10^{-1} \cite{review,web}^* \\ 
& \lambda'_{233} \lambda'_{333} & 1.2~~ 10^{-9} & 9.4~~ 10^{-10} 
& 2.0~~ 10^{-10}  & 1.7~~ 10^{-1} \cite{review}^* \\ 
\hline
& \lambda'_{311} \lambda'_{311} & 1.9~~ 10^{-3} & 2.2~~ 10^{-3} 
& 1.1~~ 10^{-2} &  1~~ 10^{-2} \cite{review}^*\\ 
& \lambda'_{312} \lambda'_{321} & 9.3~~ 10^{-5} & 1.1~~ 10^{-4} 
& 5.4~~ 10^{-4} & 1~~ 10^{-9}  \cite{review}\\ 
M_{33} 
& \lambda'_{313} \lambda'_{331} & 2.0~~ 10^{-6} & 2.4~~ 10^{-6} 
& 1.2~~ 10^{-5} & 8~~ 10^{-8}  \cite{review}\\ 
& \lambda'_{322} \lambda'_{322} & 4.6~~ 10^{-6} & 5.5~~ 10^{-6} 
& 2.7~~ 10^{-5} &  4~~ 10^{-2} \cite{review}^*\\ 
& \lambda'_{323} \lambda'_{332} & 1.0~~ 10^{-7} & 1.2~~ 10^{-7} 
& 5.8~~ 10^{-7} &  6.1~~ 10^{-2} \cite{review,web}^* \\ 
& \lambda'_{333} \lambda'_{333} & 2.2~~ 10^{-9} & 2.6~~ 10^{-9} 
& 1.3~~ 10^{-8} &  2.3~~ 10^{-1} \cite{review}^*\\ 
\hline
\end{array}
$$
%\end{center}

\newpage

%\begin{center} 
{\sf \noindent Table 2: Values ({\em not} bounds) of the
$\lambda$-type product couplings for different possible $\nu_1$
masses. In deriving our numbers, we have used the results of
ref.~\cite{fogli}. The products marked with `*' are obtained by
multiplying the upper bounds on the individual couplings.}
 $$
\begin{array}{|c|c|c|c|c|c|} 
\hline
{\rm Mass ~ matrix} & {\rm Combinations} &  m_{\nu_1}=0~{\rm eV} 
& m_{\nu_1}=0.01~{\rm eV} & m_{\nu_1}=0.1~{\rm eV} 
& {\rm Existing ~ bounds} \\
{\rm elements} & & & & & \\
\hline
& \lambda_{122} \lambda_{122} & 6.6~~ 10^{-6} & 1.1~~ 10^{-5} 
& 7.2~~ 10^{-5} & 2.5~~ 10^{-3} \cite{review}^* \\ 
M_{11} 
& \lambda_{123} \lambda_{132} & 3.9~~ 10^{-7} & 6.6~~ 10^{-7} 
& 4.3~~ 10^{-6} & 3~~ 10^{-3} \cite{review}^* \\ 
& \lambda_{133} \lambda_{133} & 2.3~~ 10^{-8} & 3.9~~ 10^{-8} 
& 2.5~~ 10^{-7} &  9~~ 10^{-6} \cite{review}^*\\ 
\hline
& \lambda_{121} \lambda_{212} & 1.6~~ 10^{-3} & 1.0~~ 10^{-3} 
& 2.0~~ 10^{-4} &  7~~ 10^{-7} \cite{review}\\ 
M_{12} = M_{21} 
& \lambda_{123} \lambda_{232} & 4.4~~ 10^{-7} & 2.9~~ 10^{-7} 
& 5.7~~ 10^{-8} & 3~~ 10^{-3} \cite{review}^* \\ 
& \lambda_{131} \lambda_{213} & 9.3~~ 10^{-5} & 6.2~~ 10^{-5} 
& 1.2~~ 10^{-5} &  3~~ 10^{-3} \cite{review}^*\\ 
& \lambda_{133} \lambda_{233} & 2.6~~ 10^{-8} & 1.7~~ 10^{-8} 
& 3.4~~ 10^{-9} & 1.8~~ 10^{-4} \cite{review}^* \\ 
\hline
& \lambda_{121} \lambda_{312} & 6.4~~ 10^{-4} & 6.6~~ 10^{-4} 
& 1.6~~ 10^{-4} &  3~~ 10^{-3} \cite{review}^*\\ 
M_{13} = M_{31} 
& \lambda_{122} \lambda_{322} & 3.1~~ 10^{-6} & 3.1~~ 10^{-6} 
& 7.4~~ 10^{-7} &  3~~ 10^{-3} \cite{review}^*\\ 
& \lambda_{131} \lambda_{313} & 3.8~~ 10^{-5} & 3.9~~ 10^{-5} 
& 9.2~~ 10^{-6} &  1.8~~ 10^{-4} \cite{review}^*\\ 
& \lambda_{132} \lambda_{323} & 1.8~~ 10^{-7} & 1.9~~ 10^{-7} 
& 4.4~~ 10^{-8} &  3.6~~ 10^{-3} \cite{review}^* \\ 
\hline
& \lambda_{211} \lambda_{211} & 4.2~~ 10^{-1} & 6.1~~ 10^{-1} 
& 3.2 &  2.5~~ 10^{-3} \cite{review}^*\\ 
M_{22} 
& \lambda_{213} \lambda_{231} & 1.2~~ 10^{-4} & 1.7~~ 10^{-4} 
& 9.1~~ 10^{-4} &  3~~ 10^{-3} \cite{review}^*\\ 
& \lambda_{233} \lambda_{233} & 3.3~~ 10^{-8} & 4.8~~ 10^{-8} 
& 2.6~~ 10^{-7} &  3.6~~ 10^{-3} \cite{review}^*\\ 
\hline
& \lambda_{211} \lambda_{311} & 3.0~~ 10^{-1} & 2.4~~ 10^{-1} 
& 5.2~~ 10^{-2} &  3~~ 10^{-3} \cite{review}^*\\ 
M_{23} = M_{32} 
& \lambda_{212} \lambda_{321} & 1.4~~ 10^{-3} & 1.1~~ 10^{-3} 
& 2.5~~ 10^{-4} &  3~~ 10^{-3} \cite{review}^*\\ 
& \lambda_{231} \lambda_{313} & 8.4~~ 10^{-5} & 6.7~~ 10^{-5} 
& 1.5~~ 10^{-5} &  1.8~~ 10^{-4} \cite{review}^*\\ 
& \lambda_{232} \lambda_{323} & 4.0~~ 10^{-7} & 3.2~~ 10^{-7} 
& 6.9~~ 10^{-8} &  3.6~~ 10^{-3} \cite{review}^* \\ 
\hline
& \lambda_{311} \lambda_{311} & 5.6~~ 10^{-1} & 6.6~~ 10^{-1} 
& 3.2  &  3.6~~ 10^{-3} \cite{review}^*\\ 
M_{33} 
& \lambda_{312} \lambda_{321} & 2.7~~ 10^{-3} & 3.2~~ 10^{-3} 
& 1.5~~ 10^{-2} &  3.6~~ 10^{-3} \cite{review}^* \\ 
& \lambda_{322} \lambda_{322} & 1.3~~ 10^{-5} & 1.5~~ 10^{-5} 
& 7.3~~ 10^{-5} &  3.6~~ 10^{-3} \cite{review}^*\\ 
\hline
\end{array}
$$
%\end{center}
\newpage
%\end{document}
\noindent
{\bf Figure 1:}
{\sf The one loop diagrams contributing to Majorana mass terms 
for the left-handed neutrinos. Figures 1(a) and 1(b) involve 
$\lambda'$- and $\lambda$-type couplings respectively.}
\begin{center}
\begin{picture}(500,600)(0,0)
\SetWidth{1.2}
\ArrowLine(20,450)(100,450)
\ArrowArc(180,450)(80,0,90)
\ArrowArc(180,450)(80,90,180)
\DashLine(100,450)(180,450){3}
\DashLine(180,450)(260,450){3}
\ArrowLine(320,450)(260,450)
%\GCirc(180,452){4}{0}
\Text(180,452)[]{$\LARGE{\boldmath{{\bullet}}}$}
\Text(180,532)[]{$\LARGE{\boldmath{{\times}}}$}
\Text(60,470)[]{$\Large{\boldmath{\nu_{_{iL}}}}$}
\Text(300,470)[]{$\Large{\boldmath{\nu_{_{iL}}}}$}
\Text(115,525)[]{$\Large{\boldmath{d_{jL}}}$}
\Text(255,525)[]{$\Large{\boldmath{d_{jR}}}$}
\Text(140,430)[]{$\Large{\boldmath{\tilde{d}_{kR}}}$}
\Text(220,430)[]{$\Large{\boldmath{\tilde{d}_{kL}}}$}
\Text(100,430)[]{$\Large{\boldmath{\lambda'_{ijk}}}$}
\Text(260,430)[]{$\Large{\boldmath{\lambda'_{ikj}}}$}
\Text(180,380)[]{\LARGE{\bf{(a)}}}

\ArrowLine(20,150)(100,150)
\ArrowArc(180,150)(80,0,90)
\ArrowArc(180,150)(80,90,180)
\DashLine(100,150)(180,150){3}
\DashLine(180,150)(260,150){3}
\ArrowLine(320,150)(260,150)
%\GCirc(180,151){4}{0}
\Text(180,151)[]{$\LARGE{\boldmath{{\bullet}}}$}
\Text(180,231)[]{$\LARGE{\boldmath{{\times}}}$}
\Text(60,170)[]{$\Large{\boldmath{\nu_{_{iL}}}}$}
\Text(300,170)[]{$\Large{\boldmath{\nu_{_{iL}}}}$}
\Text(115,225)[]{$\Large{\boldmath{e_{jL}}}$}
\Text(255,225)[]{$\Large{\boldmath{e_{jR}}}$}
\Text(140,130)[]{$\Large{\boldmath{\tilde{e}_{kR}}}$}
\Text(220,130)[]{$\Large{\boldmath{\tilde{e}_{kL}}}$}
\Text(100,130)[]{$\Large{\boldmath{\lambda_{ijk}}}$}
\Text(260,130)[]{$\Large{\boldmath{\lambda_{ikj}}}$}
\Text(180,80)[]{\LARGE{\bf{(b)}}}

\end{picture}
\end{center}
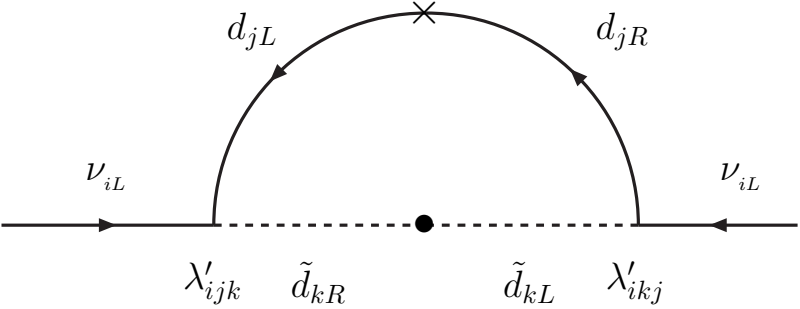
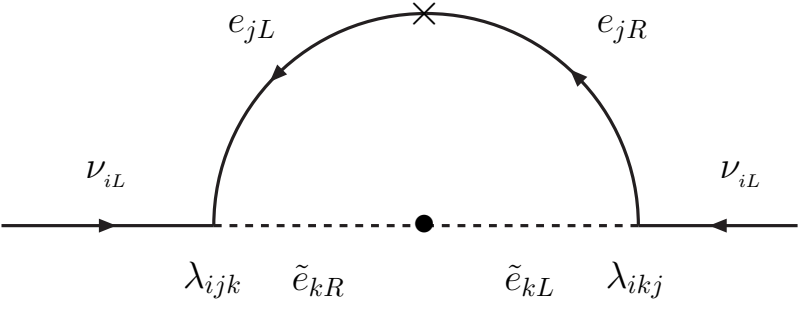
\end{document}